\documentclass[a4paper,aps,pra,showpacs,twocolumn,floatfix,superscriptaddress]{revtex4}

\usepackage{amsmath,amssymb}
\usepackage[dvips]{graphicx}

\usepackage{color}
\usepackage{tabularx}
\usepackage{graphicx}
\usepackage{dcolumn}
\usepackage{bm}

\usepackage{here}

\begin{document}


\title{Operational Magic Intensity for Sr Optical Lattice Clocks} 
\author{Ichiro Ushijima}
\affiliation{Quantum Metrology Laboratory, RIKEN, Wako, Saitama 351-0198, Japan}
\affiliation{Department of Applied Physics, Graduate School of Engineering, The University of Tokyo, Bunkyo-ku, Tokyo 113-8656, Japan}
\author{Masao Takamoto}
\affiliation{Quantum Metrology Laboratory, RIKEN, Wako, Saitama 351-0198, Japan}
\affiliation{Space-Time Engineering Research Team, RIKEN, Wako, Saitama 351-0198, Japan}

\author{Hidetoshi Katori}
\affiliation{Quantum Metrology Laboratory, RIKEN, Wako, Saitama 351-0198, Japan}
\affiliation{Department of Applied Physics, Graduate School of Engineering, The University of Tokyo, Bunkyo-ku, Tokyo 113-8656, Japan}
\affiliation{Space-Time Engineering Research Team, RIKEN, Wako, Saitama 351-0198, Japan}

\date{\today}

\begin{abstract}
We experimentally investigate the lattice-induced light shift by the electric-quadrupole ($E2$) and magnetic-dipole ($M1$) polarizabilities and the hyperpolarizability in Sr optical lattice clocks. 
Precise control of the axial  as well as the radial  motion of atoms in a one-dimensional lattice 
allows observing the $E2$-$M1$ polarizability difference.
Measured polarizabilities  determine an operational   lattice depth to be $72(2) E_R$, where the total light shift cancels to the $10^{-19}$ level, over a  lattice-intensity variation of about 30\%.
This operational trap depth and its allowable intensity range conveniently coincide with experimentally feasible operating conditions for Sr optical lattice clocks.
\end{abstract}

\pacs{06.30.Ft, 32.60.+i, 37.10.Jk, 42.62.Eh}

\keywords{Suggested keywords}

\maketitle

Recent progress of optical clocks has pushed their fractional uncertainty to the $10^{-18}$ level~\cite{Cho10,Ushijima2015, Nicholson2015,Hun16}, which opens up new applications of clocks, such as chronometric geodesy~\cite{Lisdat2016,Takano2016}, tests of fundamental constants \cite{Uza03,Hun14}, detection of dark matter~\cite{Derevianko2014}, or gravitational waves~\cite{Kol16}. 
Triggered by these advances, a future redefinition of the second by optical clocks~\cite{Targat2013,Gre16} is in range and its procedure  is being discussed~\cite{Rie18}. 

A better understanding and control of perturbations lies at the heart of the continued progress in atomic clocks. 
Isolating atoms from electromagnetic (EM) perturbations is of prime importance in designing  ion clocks~\cite{Deh82} where ions are confined nearly free from EM perturbations.
Optical lattice clocks have shown that cancellation of trap perturbation leads to stable and accurate clocks with uncertainties less than $10^{-17}$~\cite{Ushijima2015, Nicholson2015,Gre16,Bro17},
the magic frequency  aimed to equalize polarizabilities of the clock states so as to decouple the clock transition frequency from inhomogeneous trap perturbations~\cite{Katori2003}. 
Removal of perturbations  by specifying the frequency is the essence of the optical lattice clock, which is based on the fact that the frequency is a precisely measurable quantity.

This magic frequency concept, however, becomes nontrivial  for achieving inaccuracy of $10^{-18}$ because of  non-negligible contribution of the higher-order light shifts than that given by  the electric-dipole ($E1$) interaction. 
In a standing wave of light, a quarter-wavelength spatial mismatch between the $E1$ potential and the potential induced by the  electric-quadrupole ($E2$) and magnetic-dipole ($M1$)  interactions introduces an atomic-motion-dependent light shift~\cite{Taichenachev2008,Katori2009}.
In addition, the hyperpolarizability effect introduces a light shift proportional to the square of lattice  intensity~\cite{Katori2003,Brusch2006}. 
Different spatial dependence makes these light shifts difficult to eliminate.
An operational magic frequency~\cite{Katori2015} is proposed to compensate the higher order shifts by the $E1$ light shift and make the overall light shift insensitive to lattice-intensity variation around a ``magic intensity.''   
 
In order to find such an operational condition, 
precise knowledge of the higher-order polarizabilities is mandatory.
Higher-order light shifts have been investigated theoretically~\cite{Ovsiannikov2016,Porsev2018} and experimentally for Sr~\cite{Westergaard2011, Targat2013, Nicholson2015}, Yb~\cite{Barber2008,Nemitz2016,Bro17}, and Hg~\cite{Yamanaka2015}. 
Recently, the hyperpolarizability was measured for Yb to find the operational magic frequency~\cite{Bro17} with the help of a theoretical calculation of the $E2$-$M1$ polarizability. 
As for Sr, in spite of significant efforts, discrepancies between reported  polarizabilities are not yet solved. 

In this Letter, we  investigate the hyperpolarizability and the E2-M1 polarizability 
for Sr atoms in a one-dimensional (1D) lattice.
From the nonlinear intensity dependence of the light shift, we derive the hyperpolarizability. 
The $E2$-$M1$ polarizability is evaluated by measuring the light shift difference by changing the vibrational state  of atoms in the lattice.
Using the obtained polarizabilities, we  derive two distinctive operational conditions that make 
the total light shift insensitive to lattice intensity variation  at the $10^{-19}$ level.

\begin{figure}[tb]
\includegraphics[width = \linewidth,bb=10 240 400 720]{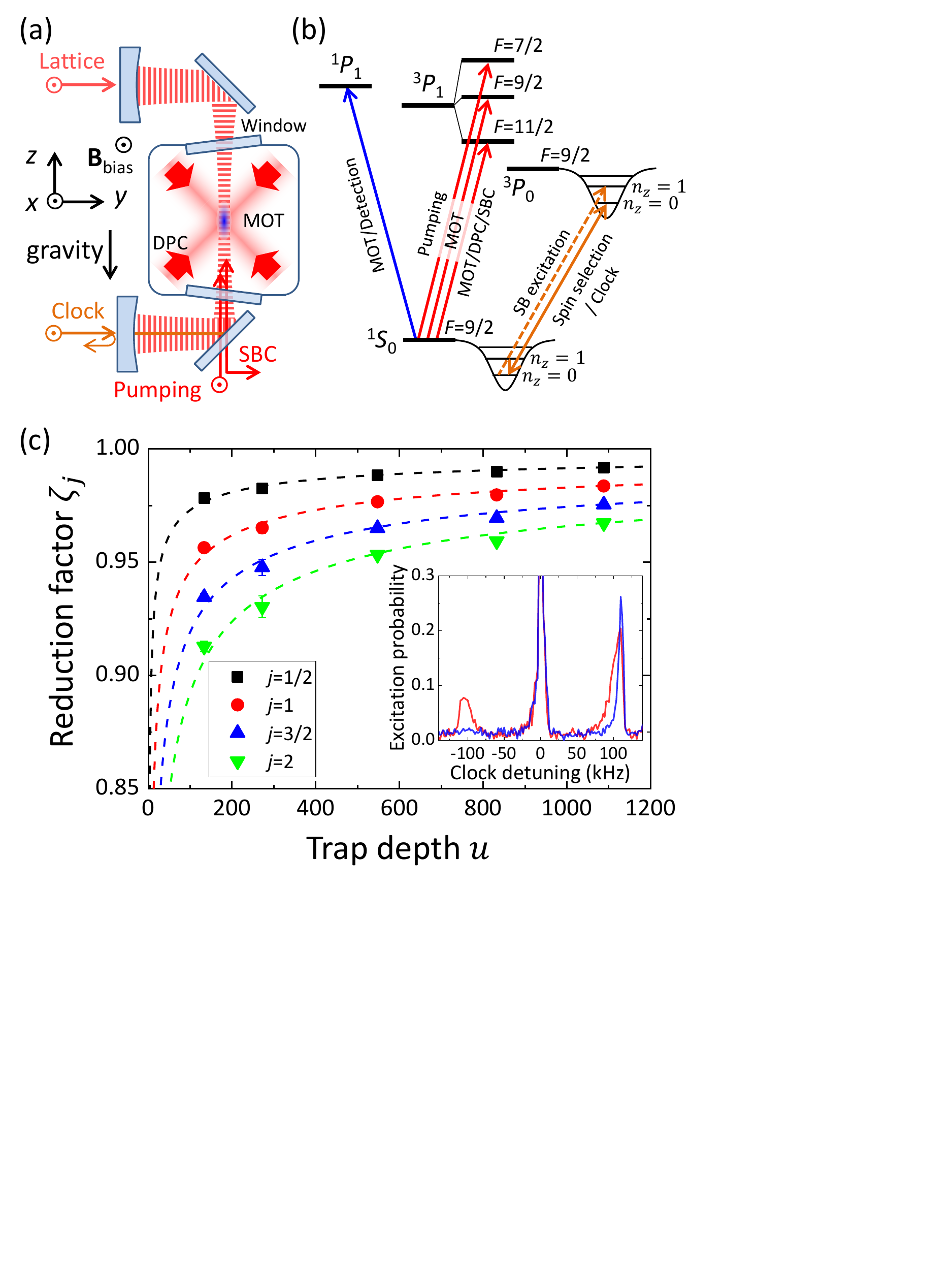} 
\caption{
(a) Experimental setup for the cavity-enhanced 1D lattice.  
After loading atoms from the magneto-optical trapping (MOT) into the lattice, we apply sideband cooling (SBC) and Doppler cooling (DPC)  on the ${^1}S_0-{^3}P_1$ transition.
(b) Energy diagram for $^{87}$Sr atoms. 
(c) Reduction factors $\zeta_j$ calculated from the radial temperature $T_r$ are shown by symbols, where colors indicate $j$ as given in the legend.
The dashed lines show estimated reduction factors   $\zeta^{\rm ad}_j (u)$  assuming the lattice depth is adiabatically varied from $u_{\rm ref}=272$ (see text). 
The blue and red lines in the inset show motional sideband spectrum on the clock transition at $u_{\rm ref}$ with and without SBC/DPC. 
}
\label{fig:Setup}
\end{figure}

The lattice-induced light shift $\nu_{\rm LS}$ is given by the  light shift difference between the ground and excited states on the clock transition. 
For a 1D optical  lattice as shown in Fig.~1(a), the  light shift  
depends on the vibrational state $n_z$ of atoms along the $z$ axis, the lattice laser intensity, 
and the  detuning $\delta_L$ of lattice laser $\nu_L=\delta_L+\nu^{E1}$ from the $E1$ magic frequency $\nu^{E1}$ that makes the $E1$ polarizabilities $\alpha^{E1}$ for  the clock states equal. 
Since the peak intensity $I_0$ of the lattice is  proportional to the trap depth $U\approx\alpha^{E1}I_0$ (by neglecting the higher-order effects of less than $10^{-6}$), we rewrite the light shift formula~\cite{Katori2015} in terms of a normalized trap depth $u=U/E_R$ with $E_R=(h \nu_L /c)^2/(2m)$ the lattice photon recoil energy as, 
\begin{align}
&h \nu_{\rm LS} (u, \delta_L, n_z)\approx  \left( \textstyle{{\partial \tilde \alpha^{E1}}\over{\partial \nu}}\delta_L -\tilde \alpha^{qm}\right) \left(n_z+\textstyle{1 \over 2} \right) u^{1/2}  \notag \\
&- \left[ \textstyle{{\partial \tilde \alpha^{E1}}\over{\partial \nu}} \delta_L+\textstyle{3 \over 2} \tilde \beta \left(n_z^2+n_z+\textstyle{1 \over 2}\right) \right] u  \notag \\
&+ 2 \tilde \beta \left(n_z+\textstyle{1 \over 2} \right) u^{3/2} - \tilde \beta \, u^2, \label{eq:LS}
\end{align}
where $\tilde \alpha^{E1}$, $\tilde \alpha^{qm}$, and $\tilde \beta$ are the difference (denoted by tildes) of $E1$ and $E2$-$M1$ polarizabilities, and hyperpolarizability on the clock transition. The conversion of these polarizabilities is summarized in the Supplemental Material~\cite{SM2017}. 
While the light shift model given in Ref.~\cite{Katori2015} takes into account the anharmonicity of the lattice trap to ${\cal O}(z^4)$  in the axial coordinate expansion, we verify that neglecting ${\cal O}(z^6)$ terms is valid for describing the light shift with low $10^{-19}$ uncertainty for Sr~\cite{SM2017}.

The  lattice intensity is nonuniform in nature, as the spatial inhomogeneity itself is the essence of an optical trap. 
As the intensity critically affects the light shift as given in Eq.~(1), precise control and evaluation of atomic distribution in the optical lattice is of particular importance. 
We consider atomic motion in the 1D lattice potential given by $\mathcal{U}(x,y,z) \approx- \alpha^{E1}I_0 {e^{ - 2({x^2} + {y^2})/{w^2}}}\cos^2 \left({2\pi z}/{\lambda_L}\right)$, where  
$I_0$, $w$, and  $\lambda_L=c/\nu_L$ are the peak intensity, the radius, and the wavelength of the lattice laser with a Gaussian profile.
The  axial and radial oscillation frequencies of atoms are  given by $\nu_z=2\sqrt{\alpha^{E1}I_0 E_R}/h$ and  ${\nu _r} = {\nu _z}{\lambda _L}/(\sqrt 2 \pi w)(\approx {\nu _z}/320$ for our experiment).
In contrast to the axial vibrational states with averaged occupation $\bar{n}_z\approx 0$ that require quantum treatment, the radial motion can be treated classically as the  vibrational states typically occupy $\bar{n}_r=k_{B}T_r/(h\nu_r)\approx 110$ with $T_r$  the radial  temperature and  $k_{B}$  the Boltzmann constant. 
Assuming a thermal distribution $\rho (x,y) = \frac{{m{(2\pi \nu_r)^2}}}{{2\pi k_B T_r}}{e^{ -\frac{1}{2} m{(2\pi \nu_r)^2}({x^2} + {y^2})/(k_B {T_r})}}$ of atoms,    
the effective   laser intensity experienced by the atoms is given by  
\begin{equation}
\overline{u^j} = \int\rho (x,y) \left( \textstyle {{\alpha^{E1}I_0 {e^{ - 2({x^2} + {y^2})/{w^2}}}}\over{E_R}}\right)^j dxdy \equiv  \zeta_j u^j,
\end{equation}
where we denote the thermal average by the bar and define a lattice-intensity reduction factor $\zeta_j(u)\approx1 - \frac{j k_{B}T_r}{u E_R}$. 
In the following, we evaluate the lattice light shifts of Eq.~(\ref{eq:LS})  by  the effective intensity $\overline{u^j} =  \zeta_j(u) u^j$.

To investigate the hyperpolarizability effect, we install a  buildup cavity with a power enhancement factor of $\approx 20$ for the 1D optical lattice oriented vertically as shown in Fig.~1(a).
The beam radius is chosen as $w \approx 60~\mu \rm{m}$ to moderate atomic collisions and allows a maximum trap depth of  $u\sim 1200$.
This cavity also works as a spatial filter to define a TEM$_{00}$ Gaussian mode. 
We use a Ti:sapphire laser at $\nu_L \approx 368~\rm{THz}$  stabilized to a reference cavity that is calibrated by a frequency comb linked to the Sr clock. 
By applying a volume Bragg grating with a bandwidth of $\sim 20~\rm{GHz}$, we suppress amplified spontaneous emission of the lattice laser and reduce the relevant light shift~\cite{Targat2013}  to  less than $10^{-19}$.

$^{87}$Sr atoms are laser cooled to $\sim 5~\mu \rm{K}$ and loaded into the lattice with its depth of $u_{\rm ref}=272$ ($u_{\rm ref}E_R/k_{B}=45\, \mu$K). This loading condition is kept constant during measurements. 
A bias magnetic field of $| \bold{B}_{\rm{bias}} | = 65~\rm{\mu T}$ is applied along the $x$ axis to define the quantization axis and to separate the Zeeman substates.
Lattice, optical pumping, and clock laser are all polarized parallel to the bias field, while that of the cooling laser is  perpendicular to the bias field so as to be decomposed into $\sigma^\pm$ components.  
Applying the $\pi$-polarized pumping laser resonant with the ${^1}S_0~(F=9/2)-{^3}P_1~(F=7/2)$ transition [see Fig.~1(b)], the atoms are optically pumped to the outermost Zeeman substates ${^1}S_0~(F=9/2,\ m_F=\pm 9/2)$ used for the clock interrogation. 
In the following, we discuss the case where we take the $m_F=9/2$ state as  the clock state. 

Simultaneously with the optical pumping, we apply Doppler cooling for the radial motion  with the $\sigma^+$ component of the cooling laser on the ${^1}S_0~(F=9/2,\ m_F=9/2)-{^3}P_1~(F=11/2,\ m_F=11/2)$ transition. 
Consequently, the radial temperature is reduced to $T_r\approx 2~\mu \rm{K}$ (correspondingly $\zeta_1(u_{\rm ref})\approx 0.96$), as measured by time-of-flight (TOF) thermometry, and the linewidth of the blue sideband on the clock transition is reduced to $\sim 8$~kHz as shown in the inset of Fig.~\ref{fig:Setup}(c).
The atoms remaining in the $m_F=-9/2$ state are heated out of the lattice by  the $\sigma^-$ component of the cooling laser. 
Subsequently, we apply sideband cooling to reduce axial vibrational states to  $\bar{n}_z<0.01$, as measured by the ratio of red and blue sidebands, using the $\sigma^+$-polarized cooling laser propagating along the lattice axis.

In order to  purify the $m_F$ state, we excite the atoms to the ${^3}P_0~(m_F=9/2)$ state with a 22-ms-long clock $\pi$ pulse so as to  resolve the Zeeman substates and  to select a single $m_F$ state. 
Atoms in the other Zeeman substates remain unexcited and are subsequently blown away by a laser pulse 
tuned to the ${^1}S_0-{^1}P_1$ transition.  
For the preparation of atoms in the  ${^3}P_0~(m_F=-9/2)$ state, we apply the similar procedure with the $\sigma^-$ component of the cooling laser.

Finally, in order to evaluate the lattice light shift dependence on the trap depth, we adiabatically  ramp up or down the lattice depth from $u_{\rm ref}$ to $u$ over 80~ms. 
Symbols in Fig.~\ref{fig:Setup}(c) show reduction factors determined by the TOF measurements, which reasonably follow  those  assuming  adiabatic temperature changes, i.e.,  $\zeta^{\rm ad}_j (u) = 1 - \frac{{1 - \zeta_j ({u_{\rm ref}})}}{{\sqrt {u/{u_{\rm ref}}} }}$ as shown by dashed lines with corresponding colors.
As the reduction factor after the adiabatic ramp is in the range of  $0.95<\zeta_1(u)<0.99$ for $150<u<1150$, we approximate $\overline{{u^j}}  \approx {({\zeta _1}u)^j}$, which is valid within 0.2\% error.
The axial vibrational number $\bar{n}_z<0.01$ is measured  unchanged after the adiabatic ramp.

\begin{figure}[t]
\includegraphics[width = \linewidth,bb=0 350 500 700]{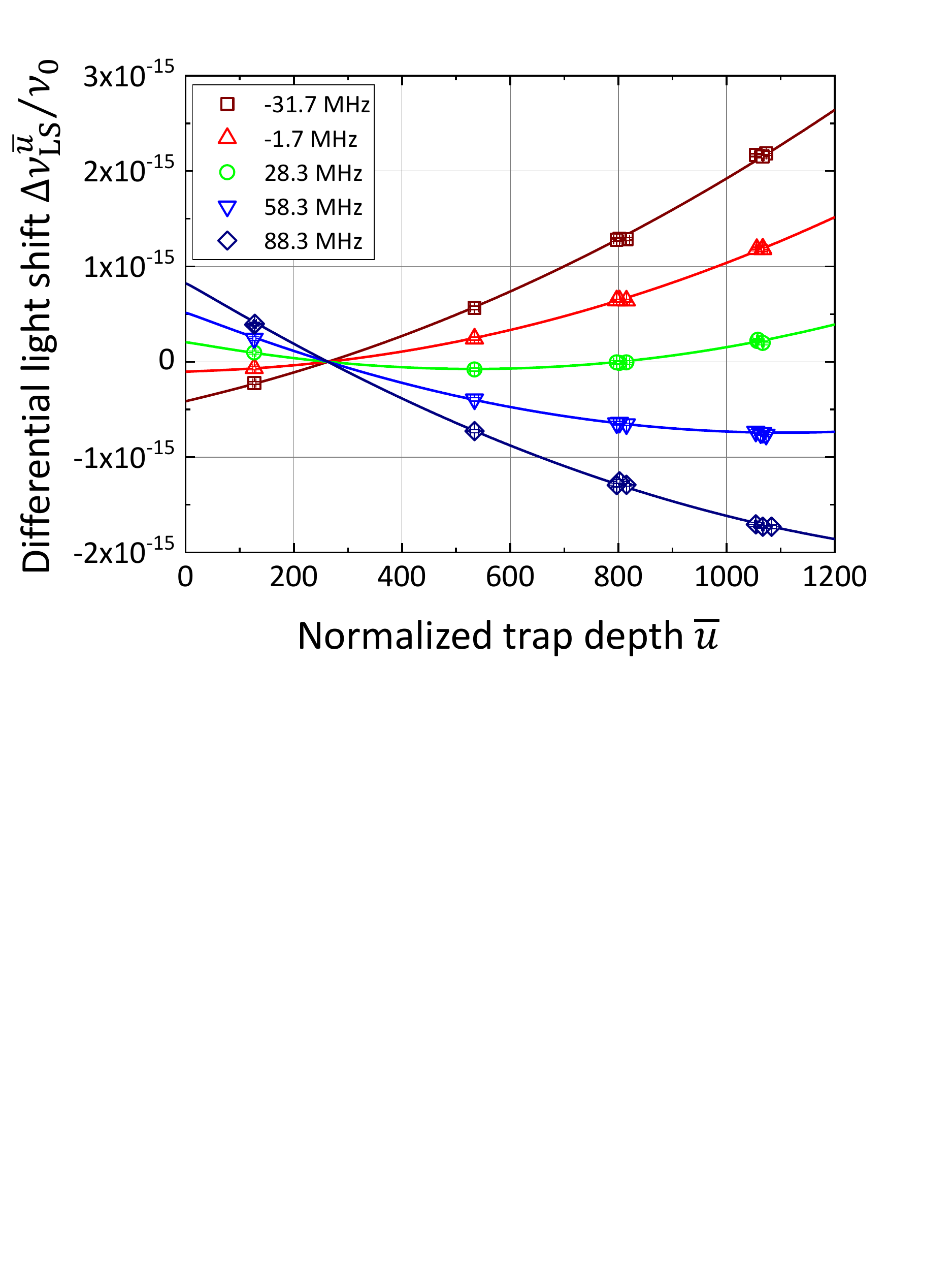} 
\caption{
Intensity-dependent  light shift $\Delta\nu_{\rm LS}^{\bar{u}}$ measured by referencing $\bar{u}_{\rm ref}=263$.
The light shifts are measured at the lattice  detunings $\delta_L$
as shown in the legend.
Error bars give 1$\sigma$ statistical uncertainties for each measurement.
The solid curves fit the measurements according  to Eq.~(\ref{eq:LS}).
}
\label{fig:Udep}
\end{figure}

We operate two Sr clocks, Sr1 and Sr2,  to evaluate the light shift:
Sr1 measures the light shift by varying the lattice depth $u$ or vibrational state $n_z$ of atoms, while Sr2 serves as a frequency anchor.
Sr1 and Sr2  simultaneously interrogate the clock transition at $\nu_0\approx 429$~THz with a common  laser to cancel out the Dick effect noise introduced by the clock laser, 
which improves the Allan deviation for the light shift measurements~\cite{Takamoto2011}.

Figure~\ref{fig:Udep} shows the  intensity-dependent light shift $\Delta\nu_{\rm LS}^{\bar{u}} = \nu_{\rm LS}(\bar{u}, \delta_L, 0)-\nu_{\rm LS}(\bar{u}_{\rm ref}, \delta_L, 0)$ as a function of the effective  depth $\bar{u}=\zeta_1(u) u$ by
taking $\bar{u}_{\rm ref} =\zeta_1(u_{\rm ref})u_{\rm ref}=263$ as a reference. 
We change the lattice laser frequency  $\nu_L$ every 30~MHz, which is measured with uncertainties less than $100~\rm{kHz}$. 
The detunings $\delta_L$  given in the legend are calculated after determining the $E1$ magic frequency $\nu^{E1}$ as described below. 
The hyperpolarizability effect introduces the nonlinear dependence for higher intensity, where we correct the density shift of  low $10^{-18}$ 
by measuring the density-dependent shift~\cite{SM2017}.

All the data in Fig.~\ref{fig:Udep} are fitted using the light shift model given in Eq.~(\ref{eq:LS}), where we take $\nu^{E1}$, $\frac{\partial \tilde \alpha^{E1}}{\partial \nu}$, and $\tilde \beta$ as free parameters.
As $\tilde \alpha^{qm}$ scarcely contributes to  this fitting, we conduct another measurement to determine $\tilde \alpha^{qm}$   and apply the results to this fitting. We repeat these two fittings until the fitting parameters converge.
Finally, the solid fitting  curves  determine 
$\nu^{E1}=368~554~465.1(1.0)~\rm{MHz}$, 
$(\partial \tilde \alpha^{E1} / \partial \nu)/h=1.735(13)\times10^{-11}$, and 
$\tilde \beta /h =-0.461(14)~\rm{\mu Hz}$.

As the light shift arising from the multipolar polarizability $\tilde \alpha^{qm}$ is sensitive to the vibrational states $n_z$~\cite{Katori2009}, 
we measure the differential light shift between $n_z=1$ and $n_z=0$ 
vibrational states given by 
\begin{align}
& h\Delta \nu_{\rm LS}^{\rm{vib}}(u, \delta_L) \notag \\
&= h [\nu_{\rm LS}(u, \delta_L, 1) - \nu_{\rm LS}(u, \delta_L, 0)] \notag \\
&=  \left( \textstyle{{\partial \tilde \alpha^{E1}}\over{\partial \nu}}\delta_L-\tilde \alpha^{qm}\right) u^{1/2} +\tilde \beta u \left(2 u^{1/2}-3\right). \label{eq:MultiPol}
\end{align}
This eliminates the otherwise dominating contributions from $\tilde\alpha^{E1}$  and $\tilde \beta$,  and allows extracting $\tilde \alpha^{qm}$.

\begin{figure}[t]
\includegraphics[width = \linewidth,bb=0 350 500 700]{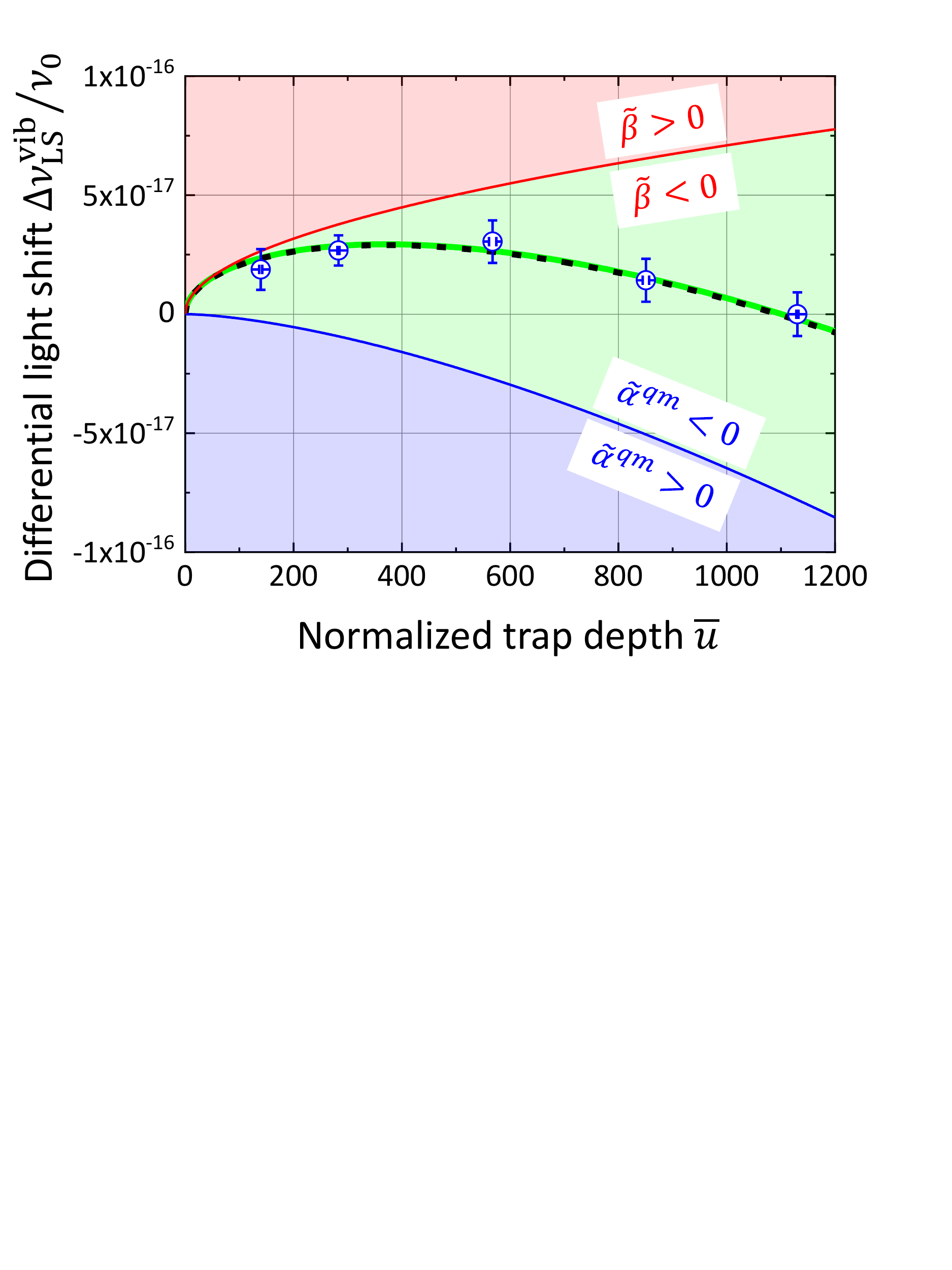} 
\caption{
Evaluation of the multipolar polarizability from the light shift difference between $n_z=1$ and $n_z=0$ 
measured at $\delta_L=0.4~\rm{MHz}$ shown by empty circles.
Assuming $\beta$ derived from Fig.~2, regions with $\tilde \alpha^{qm}< 0\,(> 0)$ are displayed by upper red-green (lower blue) area. 
Empty circles fall on the upper region, indicating $\tilde \alpha^{qm}<0$.
By taking $\tilde \alpha^{qm}$ in Eq.~(\ref{eq:MultiPol}) as a free parameter, the fitting  determines  $\tilde \alpha^{qm}/h = -0.962(40)~\rm{mHz}$ as shown by a green line.
}
\label{fig:multipolar}
\end{figure}

For this measurement, we excite the atoms to the $n_z=0$ or $1$ vibrational state in the ${^3}P_0~(m_F=9/2)$ state by applying a rapid adiabatic passage (RAP)~\cite{Melinger1992} by frequency sweeping the $\pi$-polarized clock laser across the carrier and blue sideband in 6 ms.
The Rabi frequency of the clock laser is about 50~kHz (10~kHz) for the carrier (the blue sideband). 
This RAP allows transferring more than 90\% of the atoms to the desired vibrational states. 
The atoms remaining in the ground state are heated out of the trap by driving the ${^1}S_0-{^1}P_1$ transition.

Figure~\ref{fig:multipolar} shows the differential light shift  $\Delta\nu_{\rm LS}^{\rm{vib}} (\bar{u}, \delta_L)$ measured for the lattice detuning $\delta_L=0.4~\rm{MHz}$.
A green line fits the measurements by taking $\tilde \alpha^{qm}$ as a free parameter,
while $\tilde \beta$, $\nu^{E1}$, and $\partial \tilde \alpha^{E1}/\partial \nu$ are fixed with the values obtained with the data in Fig.~2.
The updated result of $\tilde \alpha^{qm}$ is recursively used for deriving the hyperpolarizability.
We determine the differential multipolar polarizability to be $\tilde \alpha^{qm}/h = -0.962(40)~\rm{mHz}$.
The black dashed line shows  $\Delta\nu_{\rm LS}^{\rm{vib}}(\bar{u},0)$ at the $E1$ magic frequency $\nu^{E1}$.
By setting $\tilde \beta=0$ and $\tilde \alpha^{qm}=0$, we obtain red and blue lines, which indicate that  $\Delta\nu_{\rm LS}^{\rm{vib}}(\bar{u},0)$ is mainly determined by the multipolar polarizability for   $\bar{u} < 200$ and the hyperpolarizability starts to contribute for higher intensity. 
Note that the two lines divide the plot into 3 sections indicated by different colors depending on the signs of these polarizabilities. 

\begin{figure}[t]
\includegraphics[width = \linewidth,bb=0 350 500 700]{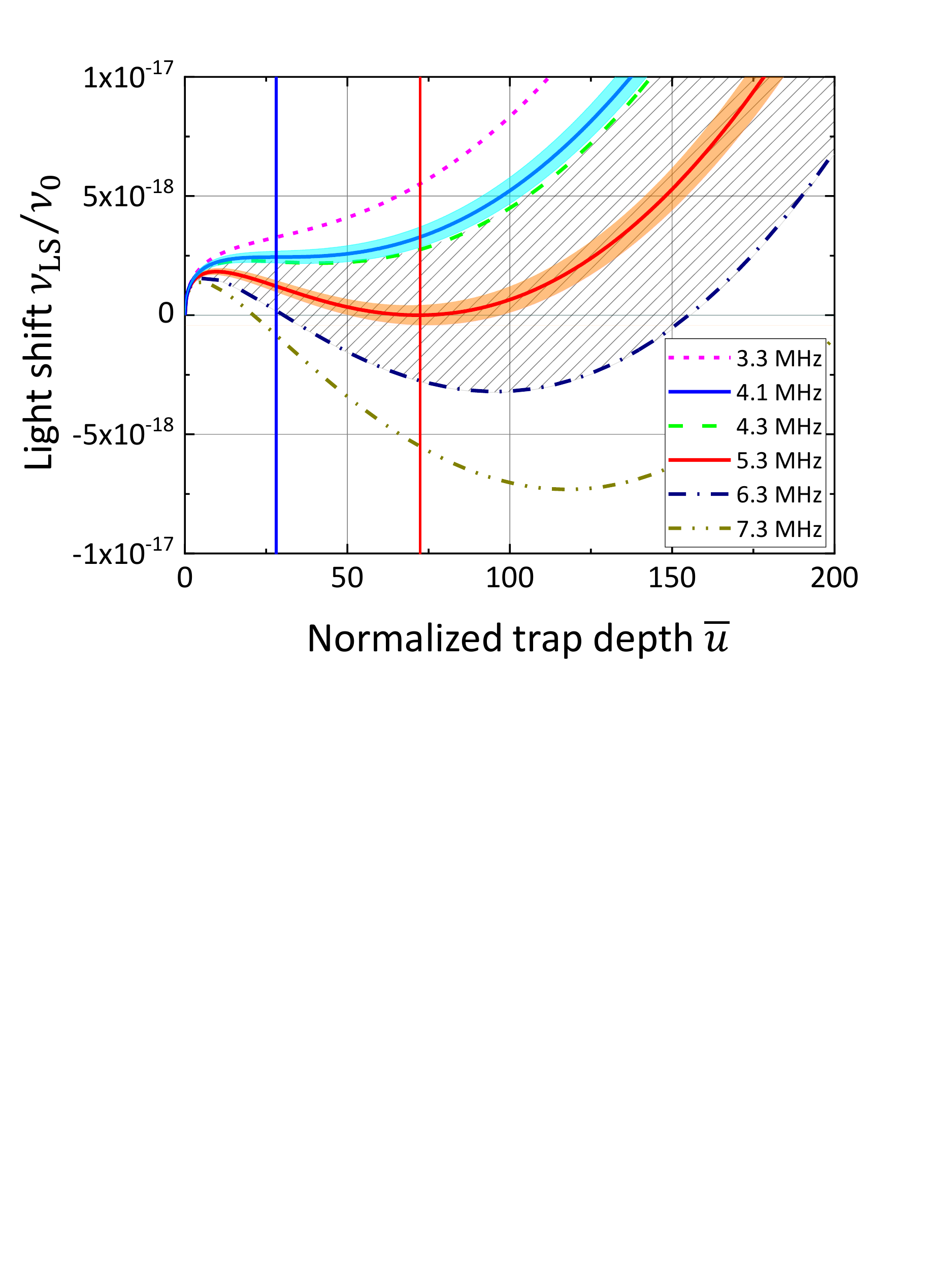} 
\caption{
Lattice light shift near the operational magic conditions for different detunings $\delta_L$.
The red and blue curves show the light shifts for two operational magic frequencies (i) $\delta_L=5.3(2)~\rm{MHz}$ to make the light shift zero at $\bar{u}^{\rm{op}}=72(2)$ (red vertical line), and (ii) $\delta_L=4.1(1)~\rm{MHz}$ to use an inflection point at $\bar{u}^{\rm{op}}=28(1)$ (blue vertical line).
Orange (sky-blue) shaded area indicates an uncertainty given by the measured polarizabilities and a hatched area indicates that given by the $E1$ magic frequency for case (i).
}
\label{fig:LSsummary}
\end{figure}

The lattice-induced light shifts $\nu_{\rm LS} (\bar{u}, \delta_L,0)$ predicted by the obtained polarizabilities are shown in Fig.~\ref{fig:LSsummary}. 
In addition to making the light shift insensitive to the trap depth $\bar{u}$, i.e., ${{\textstyle{{\partial {\nu _{{\rm{LS}}}}} \over {\partial u}} }}|_{u=\bar{u}^{\rm op}} = 0$, the Sr clock transition offers two distinctive operational  conditions $(\bar{u}^{\rm op}, \delta_L^{\rm op})$,  as it has the same sign for $\tilde \beta$ and $\tilde \alpha^{qm}$~\cite{SM2017} as indicated by the green area in Fig.~3:  
(i) by taking $\delta_L^{\rm op}=5.3(2)~\rm{MHz}$ and $\bar{u}^{\rm{op}}=72(2)$, the total  light shift can be reduced to less than $1\times 10^{-19}$ over the trap depth  $60<u<83$ as indicated by a red line. 
Alternatively, (ii)  by taking $\delta_L^{\rm op}=4.1(1)~\rm{MHz}$ and $\bar{u}^{\rm{op}}=28(1)$, an inflection point determined by 
${{\textstyle{{\partial^2 {\nu _{{\rm{LS}}}}} \over {\partial u^2}} }}|_{u=\bar{u}^{\rm op}} = 0$ offers  the light shift variation less than  $1\times10^{-19}$ over the trap depth $17<u<43$ as shown by a blue line.
Orange and sky-blue shaded areas indicate the uncertainties of $4\times 10^{-19}$ and $2\times 10^{-19}$ given by those of measured polarizabilities.
The $E1$ magic frequency uncertainty of 1.0 MHz for the present measurements, including the tensor-shift contribution as discussed in the  Supplemental Material~\cite{SM2017}, gives an overall light-shift uncertainty $3\times 10^{-18}$ at $\bar{u}^{\rm{op}}=72$ (hatched area) and $1\times 10^{-18}$ at $\bar{u}^{\rm{op}}=28$, which can be  reduced by improving the statistics of the clock measurements.
For the lattice depth of $72E_R$ and $28E_R$, the off-resonant lattice-photon scattering rate~\cite{Dorscher2018}, including Raman scattering in the $^3P_0$ state and Rayleigh scattering, is estimated to be $0.1$ and $0.04\,{\rm s}^{-1}$, allowing a sufficient clock interrogation time over multiple seconds.

\begin{figure}[t]
\includegraphics[width = \linewidth,bb=20 550 550 720]{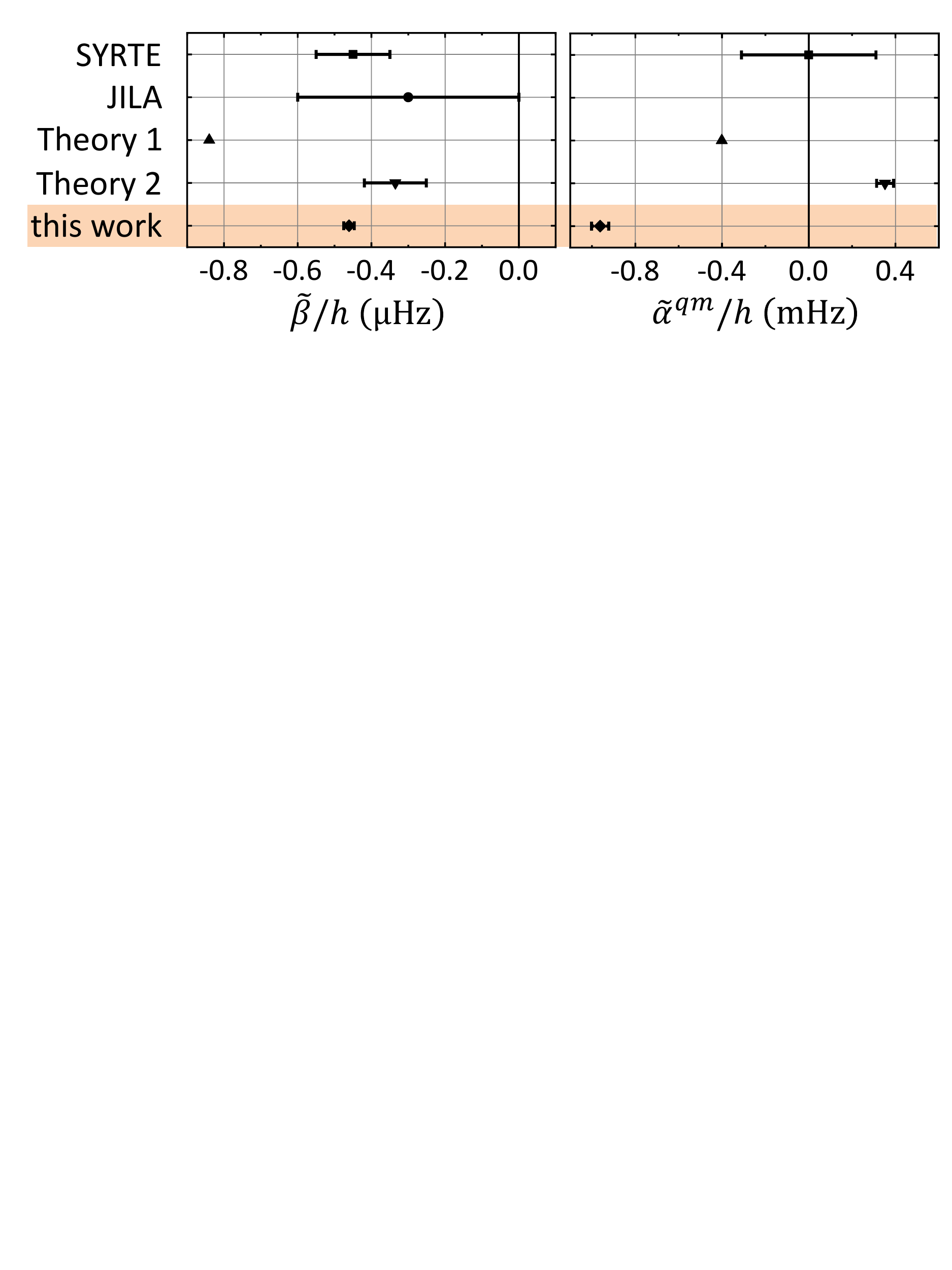} 
\caption{
Summary of differential hyperpolarizability $\tilde \beta$ and multipolar polarizability $\tilde \alpha^{qm}$ on the clock transition reported in previous works SYRTE~\cite{Westergaard2011,Targat2013}, JILA~\cite{Nicholson2015}, theory~1 (error bars not available)~\cite{Ovsiannikov2016}, theory~2~\cite{Porsev2018}, and this work.
}
\label{fig:comparison}
\end{figure}

Figure~\ref{fig:comparison} summarizes reported  polarizabilities for the $^1S_0-{}^3P_0$ clock transition of Sr.
The hyperpolarizability $\tilde \beta$ determined in this work agrees with the previous results \cite{Westergaard2011,Nicholson2015} within their uncertainties and is close to a recent theory \cite{Porsev2018}.
Our multipolar polarizability $\tilde \alpha^{qm}$  deviates from the previous experiment \cite{Westergaard2011} that indicates zero within the uncertainty, and  from two theories \cite{Ovsiannikov2016,Porsev2018} that give opposite signs with each other.

In summary, we have determined the differential multipolar $(\tilde \alpha^{qm})$ and hyper $(\tilde \beta)$ polarizabilities for Sr optical lattice clocks by precisely controlling the atomic motion. 
These  polarizabilities predict two distinctive operational conditions: the lattice depth and frequency $\delta_L^{\rm op}$  of $(72E_R,  5.3\,{\rm MHz})$ allows canceling out the lattice light shift and $(28E_R,  4.1\,{\rm MHz})$ allows using the inflection point, both of which  coincide with typical operating conditions for Sr  clocks~\cite{Ushijima2015,Dorscher2018}. 
These operational lattice depths are conveniently described  by magic  sideband frequencies of $\nu_z^{\rm op}=59(1)/\sqrt{\zeta_1}$~kHz and $29(1)/\sqrt{\zeta_1}$~kHz for the axial motion, respectively, with $\zeta_1$ the intensity reduction factor to be measured. 
A narrow-line cooling~\cite{Kat99}  allows $\zeta_1\approx 0.91$ or better, which well meets the predicted lattice intensity tolerance of more than 30\% around the magic intensity.
Combined with cryogenic clocks that  reduce the blackbody radiation shift~\cite{Ushijima2015}, the clock uncertainty at the level of $10^{-19}$ falls  within the scope.

We thank M. Das for his contribution to the experiments in the early stage, N. Ohmae for the operation of the comb, and N. Nemitz for careful readings of the manuscript and valuable comments. 
This work is supported by JST ERATO Grant No. JPMJER1002 10102832 (Japan), by JSPS Grant-in-Aid for Specially Promoted Research Grant No. JP16H06284, and by the Photon Frontier Network Program of the Ministry of Education, Culture, Sports, Science and Technology, Japan (MEXT).


\end{document}


\title{Supplemental Material to\\ Operational Magic Intensity for Sr Optical Lattice Clocks}

\affiliation{Quantum Metrology Laboratory, RIKEN, Wako, Saitama 351-0198, Japan}
\affiliation{Department of Applied Physics, Graduate School of Engineering, The University of Tokyo, Bunkyo-ku, Tokyo 113-8656, Japan}
\affiliation{Space-Time Engineering Research Team, RIKEN, Wako, Saitama 351-0198, Japan}

\author{Ichiro Ushijima}
\affiliation{Quantum Metrology Laboratory, RIKEN, Wako, Saitama 351-0198, Japan}
\affiliation{Department of Applied Physics, Graduate School of Engineering, The University of Tokyo, Bunkyo-ku, Tokyo 113-8656, Japan}
\author{Masao Takamoto}
\affiliation{Quantum Metrology Laboratory, RIKEN, Wako, Saitama 351-0198, Japan}
\affiliation{Space-Time Engineering Research Team, RIKEN, Wako, Saitama 351-0198, Japan}
\author{Hidetoshi Katori}
\affiliation{Quantum Metrology Laboratory, RIKEN, Wako, Saitama 351-0198, Japan}
\affiliation{Department of Applied Physics, Graduate School of Engineering, The University of Tokyo, Bunkyo-ku, Tokyo 113-8656, Japan}
\affiliation{Space-Time Engineering Research Team, RIKEN, Wako, Saitama 351-0198, Japan}

\date{\today}
\pacs{06.30.Ft, 32.60.+i, 37.10.Jk, 42.62.Eh} 
\maketitle 

\section*{Determination of electric dipole polarizability}
By measuring the potential depth $U$ and  the peak intensity $I_0$ of lattice laser  inside the optical cavity, we evaluate the electric-dipole polarizability $\alpha^{\rm E1}\approx U/I_0$ of the ${^{1}S_{0}}$ and ${{^{3}P_{0}}}$ states  near the E1 magic frequency.
The potential depth of the lattice $U={h^2 \nu_z^2}/({4 E_R})$ is estimated by the axial trap frequency $\nu _z$, which is determined by the sideband spectroscopy~\cite{Blatt2009}, with $E_R$  the lattice photon recoil energy.

Our lattice cavity consists of  two curved mirrors with a radius of curvature of 150 mm and 200 mm separated by about 350 mm. 
The waist size $w= \sqrt{2}(\nu _{z}/\nu _{r})/k_{L}$ of the lattice is determined by measuring the ratio $\nu _{z}/\nu _{r}$ of axial and radial trap frequencies with $k_{L}=2\pi \nu_L/c$ the wave number of the lattice laser.
By slightly tilting the clock laser in respect to the axis of the 1D optical lattice, we measure sidebands for the radial motion to determine $\nu_r=246(19)~\rm{Hz}$ for $\nu_z=79~\rm{kHz}$, which indicates  $w = 58.9(4.8)~\rm{\mu m}$.

We estimate an intra-cavity power from the reflected power $P_{\rm{r}}$ by a view-port that is slightly tilted against the lattice laser as shown in Fig.~1(a) in the main text, which is  $P_{\rm{r}}=10.4(5)$ mW for $\nu _z =166$ kHz. 
The reflectance of the window is independently measured to be $R_{\rm{w}}=0.0052(2)$.
We therefore estimate the intra-cavity power $P = P_{\rm{r}}/R_{\rm{w}} =  2.00(12)~\rm{W}$.
The peak intensity of the lattice laser $I_0=4 \frac{2P}{ \pi w^2}$ is calculated to be $I_0 = 147(9)~\rm{kW/{cm}^2}$, where a factor 4 accounts for the intensity at the anti-node of the standing wave.
By combining these measurements we determine the electric-dipole polarizability to be $\alpha^{\rm E1}/h =54.1(4.3)~\rm{kHz/(kW/{cm}^2)}$.

\section*{Higher order anharmonicity of the lattice potential and conversion of notations}
Ref.~\cite{Katori2015} takes into account the anharmonicity of the lattice trap to the 4th order.
However, for a shallow lattice potential or for the higher vibrational states $n>1$, it is not evident whether this approximation is applicable.
We therefore  calculate the contribution of anharmonicity in the 6th order. In addition, we describe the relation between slightly different notations used in Ref.~\cite{Katori2015} and in this manuscript. The new notations used in this manuscript are suitable for characterizing the lattice  in terms of frequency instead of the lattice intensity that  is difficult to measure precisely.  

Assuming the light shifts arising from the E1 interaction
$U^{\rm E1}_{g(e)}(I)=-\alpha_{g(e)} ^{\rm E1}I \cos^2(k_L z)$, 
the E2-M1 interaction $U^{qm}_{g(e)}(I)=-\alpha_{g(e)} ^{qm}I \sin^2(k_L z)$, 
and  the hyperpolarizability effect $U^{\rm{Hyp}}_{g(e)}(I)=-\beta_{g(e)} I^2 \cos^4(k_L z)$ for the ground (excited) state, denoted by the subscript $g$ $(e)$, of the clock transition with $I$ the laser intensity, the total light shift  is given by~\cite{Katori2015} 
\begin{align}
&U_{g(e)}(I)=U^{\rm E1}_{g(e)}(I)+U^{qm}_{g(e)}(I)+U^{\rm{Hyp}}_{g(e)}(I) \nonumber \\
&=-\alpha_{g(e)} ^{\rm E1}I \cos^2k_L z -\alpha_{g(e)} ^{qm}I \sin^2k_L z-\beta_{g(e)} I^2 \cos^4k_L z.
\end{align}
We Taylor-expand $U_{g(e)}(I)$ to the 6th order in $z$ as
\begin{align}
&U_{g(e)}(I) \sim -\alpha _{g(e)} ^{\rm E1}I-\beta _{g(e)} I^2 \nonumber \\
&+(k_L z)^2 \left[(\alpha _{g(e)}^{\rm E1} - \alpha _{g(e)}^{qm})I +2\beta _{g(e)}I^2 \right] \nonumber \\
&-\frac{1}{3}(k_L z)^4 \left[(\alpha _{g(e)}^{\rm E1} - \alpha _{g(e)}^{qm})I +5\beta _{g(e)}I^2 \right] \nonumber \\ 
&+\frac{2}{45}(k_L z)^6 \left[(\alpha _{g(e)}^{\rm E1} - \alpha _{g(e)}^{qm})I +17\beta _{g(e)}I^2 \right]. \nonumber \\
\end{align}
The light shift $h\nu_{\rm LS}(I,n)$ on the clock transition for the $n$-th vibrational state $ |n\rangle$ is calculated by the light shift difference $\Delta U(I)=U_{e}(I)-U_{g}(I)$ as,
\begin{equation}
h\nu_{\rm LS}(I,n)\approx\langle n|\Delta U(I) |n\rangle=h\nu^{(4)}_{\rm LS}(I,n)+h\nu^{(6)}_{\rm LS}(I,n),
\end{equation}
where $\nu^{(4)}_{\rm LS}$ term represents the Taylor series up to $z^4$ and $\nu^{(6)}_{\rm LS}$ gives the $z^6$ term.
Denoting the differential electric-dipole polarizability 
\[\Delta \alpha^{\rm E1}=\alpha _{e}^{\rm E1}-\alpha _{g}^{\rm E1},\] 
E2-M1 polarizability 
\[\Delta \alpha ^{qm}=\alpha _{e}^{qm}-\alpha _{g}^{qm},\] 
and hyperpolarizability 
\[\Delta \beta =\beta _{e}-\beta _{g},\] 
corresponding light shifts are given by 
\begin{align}
&h\nu^{(4)}_{\rm LS}= -\left[\Delta \alpha ^{\rm E1}+\frac{3}{2}\frac{E_R}{\alpha ^{\rm E1}}\Delta \beta\left(n^2+n+\frac{1}{2}\right)\right]I \nonumber \\
&+\sqrt{\frac{E_R}{\alpha ^{\rm E1}}}(\Delta \alpha ^{\rm E1}-\Delta \alpha ^{qm})\left(n+\frac{1}{2}\right) I^{1/2} \nonumber \\
&+2\sqrt{\frac{E_R}{\alpha ^{\rm E1}}}\Delta \beta \left(n+\frac{1}{2} \right)I^{3/2} - \Delta \beta I^2, \nonumber \\
\label{eq:4th}
\end{align}
which is identical to the formula given in Ref.~\cite{Katori2015}, and  
\begin{align}
&h\nu^{(6)}_{\rm LS}=\frac{14}{9}\left(\frac{E_R}{\alpha ^{\rm E1}}\right)^{3/2}\Delta \beta \left(n^3+\frac{3}{2}n^2+2n+\frac{3}{4}\right) I^{1/2} \nonumber \\
&-\frac{1}{18}\left(\frac{E_R}{\alpha ^{\rm E1}}\right)^{3/2}(\Delta \alpha ^{\rm E1}-\Delta \alpha ^{qm})\nonumber \\
&\times \left(n^3+\frac{3}{2}n^2+2n+\frac{3}{4}\right) I^{-1/2},
\label{eq:6th}
\end{align}
gives the $z^6$ order correction.

\begin{figure}[t]
\includegraphics[width = \linewidth, bb=0 150 530 750]{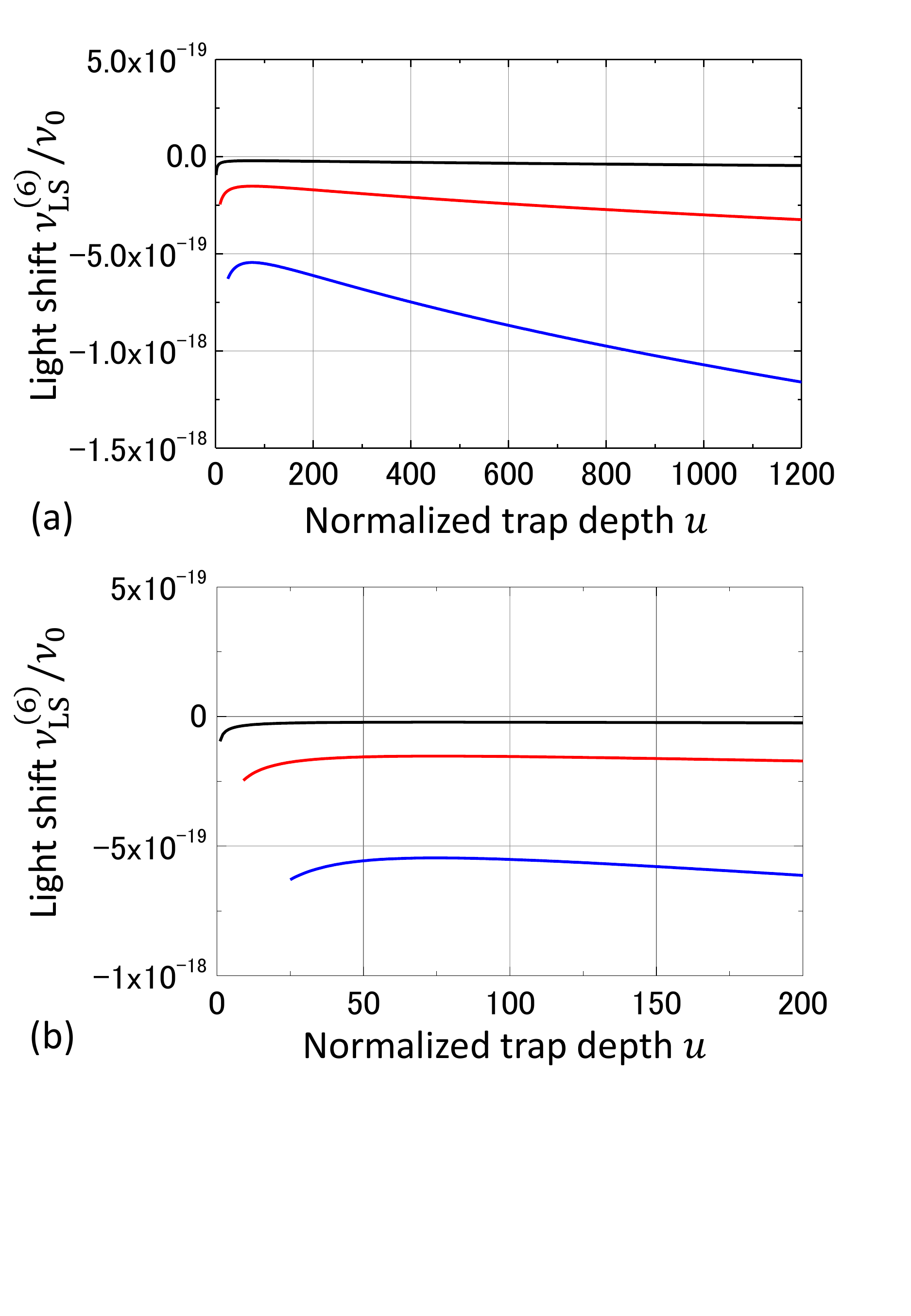}
\caption{
Intensity dependent light shift $h\nu^{(6)}_{\rm LS}/\nu_0$ at the E1 magic frequency. 
The black, red, and blue lines show the calculations  of light shifts for $n=0$, 1, and 2 for the trap depth in the range of  $0<u<1200$ and $0<u<200$ displayed in (a) and (b), respectively.
}
\label{fig0:6th}
\end{figure}

We apply new notations in the unit of $E_R/\alpha^{\rm E1}$ (given in the left hand side), which are used in the main text and in the following discussion,
 \begin{align}
&\tilde \alpha^{\rm E1}\equiv \Delta \alpha^{\rm E1}E_R/\alpha^{\rm E1},\nonumber \\
&\tilde \alpha^{qm}\equiv \Delta \alpha^{qm}E_R/\alpha^{\rm E1},\nonumber \\
&\tilde \beta\equiv \Delta \beta\left(E_R/\alpha^{\rm E1}\right)^2,\nonumber \\
&u\equiv I(E_R/\alpha^{\rm E1})^{-1},
\end{align}
which allows  describing measurements by frequencies instead of intensities.
Using these notations, Eq.~(\ref{eq:4th}) becomes identical to Eq.~(1)    in the main text. Eq.~(\ref{eq:6th}) becomes
\begin{align}
&h\nu^{(6)}_{\rm LS}(u)=\frac{14}{9} \left(n^3+\frac{3}{2}n^2+2n+\frac{3}{4}\right)\tilde \beta u^{1/2} \nonumber \\
&-\frac{1}{18}\left(n^3+\frac{3}{2}n^2+2n+\frac{3}{4}\right)(\tilde \alpha ^{\rm E1}-\tilde \alpha ^{qm}) u^{-1/2}.
\label{eq:6th1}
\end{align}

Figure \ref{fig0:6th} shows the fractional contribution of  $h\nu^{(6)}_{\rm LS}/\nu_0$ at the E1 magic frequency, where $\tilde \alpha ^{\rm E1}=0$ holds,  as a function of the normalized lattice depth $u$, with $\nu_0\approx429$~THz the clock transition frequency. 
The black, red, and blue curves correspond to the vibrational states $n=0$, 1, and 2, respectively.
In the range of  $20<u<1200$ and for $n=0, 1$, where we investigate the light shift, $|h\nu^{(6)}_{\rm LS}/\nu_0|<3.3\times 10^{-19}$ validates that 6th order contribution is sufficiently small compared to our measurement uncertainty.

\section*{Uncertainty of the E1 magic frequency and the hyperpolarizability}
For the precise determination of $\nu^{\rm E1}$ and $\tilde \beta$, we have investigated the tensor and vector light shifts. 
The tensor shift may introduce a frequency offset to the E1 magic frequency \cite{Takamoto2006,Westergaard2011,Targat2013}.
In our lattice configuration, the lattice laser polarization is set parallel to the bias magnetic field $\bold{B}_{\rm{bias}}$ with the angle uncertainty  less than $30~\rm{mrad}$, corresponding to $0.3~\rm{MHz}$ uncertainty of  $\nu^{\rm E1}$.
The vector light shift is used to estimate the degree of ellipticity $\xi_L\approx 2.6 \times 10^{-4}$ of the lattice laser polarization introduced by the birefringence of the vacuum chamber windows [see Fig.~1(a) in the main text].
Taking the ellipticity-dependent hyperpolarizability~\cite{Ovsiannikov2016} of $\tilde{\beta}^{c}\approx1.5\tilde{\beta}^{l}$ for the circular $(\tilde{\beta}^{c})$ and linear  $(\tilde{\beta}^{l})$  polarizations into account, the fractional uncertainties for the hyperpolarizability originating from the birefringence is less than $10^{-7}$.

\section*{Vector light shift}
\begin{figure}[t]
\includegraphics[width =\linewidth,bb=30 400 530 750]{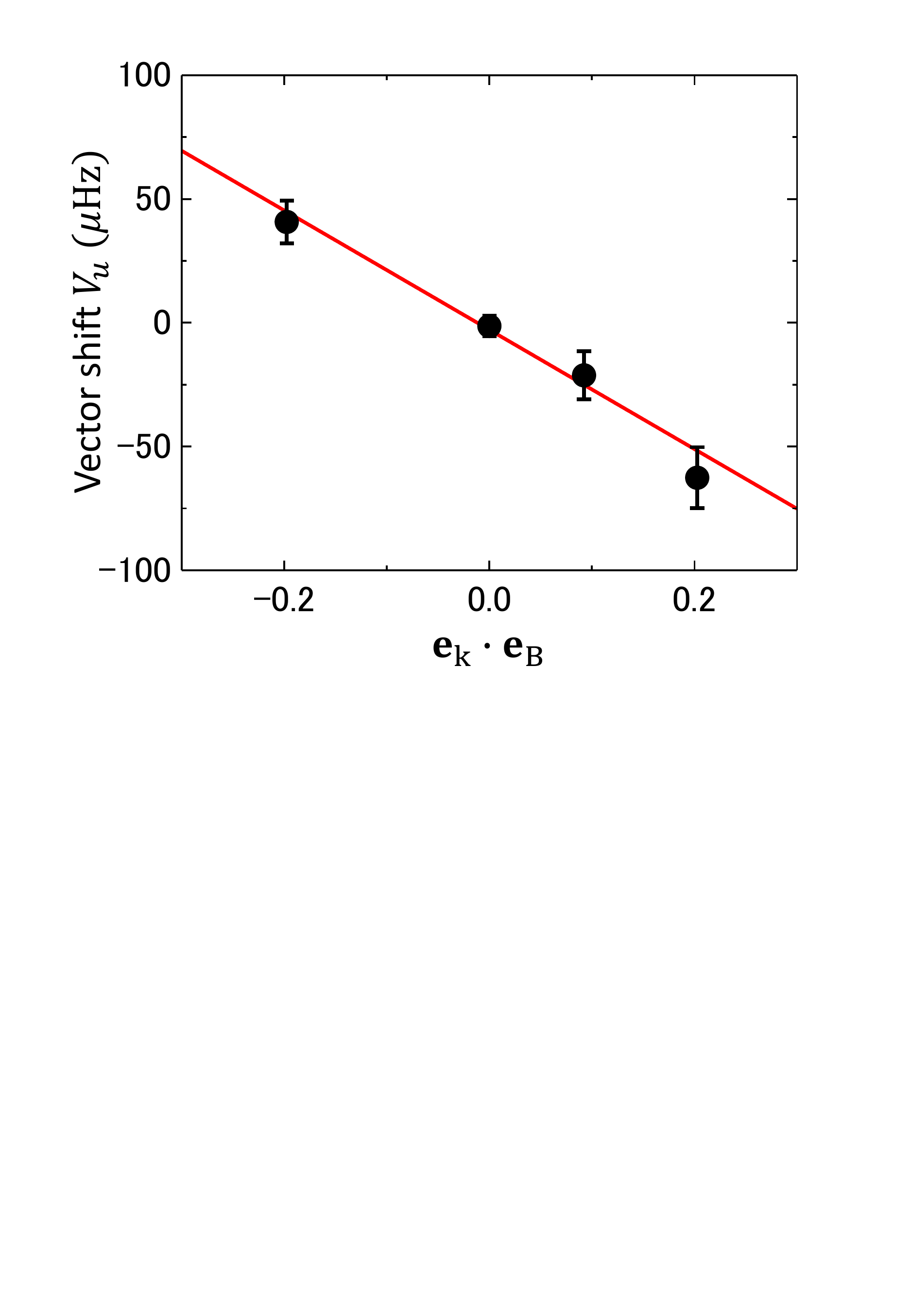}
\caption{
The dependence of the vector light shift on the angle of the lattice wave vector and the quantization axis. 
The red line shows the result of a linear fitting.
}
\label{vector}
\end{figure}
The E1 light shift $\nu _{\rm{LS}}^{\rm E1}$ is given by the sum of the scalar $S_u$, vector $V_u$, and tensor $T_u$ light shifts that are induced by the differences of scalar $\tilde \alpha^{\rm E1}$, vector ${\tilde \kappa}^v$, and tensor ${\tilde \kappa}^t$, polarizabilities between the clock states,
\begin{align} 
\label{E1}
&h \nu _{\rm{LS}}^{\rm E1} (u,\delta _{\it{L}}, m_{F})\nonumber \\
&\equiv \left(S_u+V_u+T_u\right)u\nonumber \\
&= \left(-\frac{\partial \tilde \alpha ^{\rm E1}}{\partial \nu _{\it{L}}}\delta_{\it{L}}
+\tilde \kappa ^{v} m_{\it{F}}\xi _{L} {\bf e}_{\rm k} \cdot {\bf e}_{\rm B}+\tilde \kappa ^{t}b \right)u,
\end{align} 
where $b = (3 |{\bf e}_{\epsilon} \cdot {\bf e}_{\rm B}|^{2}-1)[3m_{F}^{2}-F(F+1)]$, ${\bf e}_{\rm k}$ is the unit vector along the lattice wave vector that is parallel to ${\bf e}_z$, ${\bf e}_{\rm B}={\bf B}_{\rm tot}/|{\bf B}_{\rm tot}|$ is  the quantization axis, and 
${\bf e}_{\epsilon}$ is the complex polarization vector of the lattice laser.

We measure frequency differences between two outermost Zeeman components ($m_{F}=\pm 9/2$) of $^{87}$Sr clock transitions as 
$\Delta \nu_{\rm vec}({u})=\nu_{+9/2} ({u})- \nu_{-9/2}({u})$. 
In order to remove the first order Zeeman shift, we measure the intensity-dependent $(\Bar{u})$ light shift and determine the vector light shift as
\begin{equation}
V_u=\frac{\Delta \nu_{\rm vec}(\bar{u})-\Delta \nu_{\rm vec}(\bar{u}_{\rm ref})}{2(\bar{u}-\bar{u}_{\rm ref})},
\end{equation}
where we vary effective trap depth $\bar{u}$ and take $\bar{u}_{\rm ref}=280$.

Figure~\ref{vector} shows the normalized vector light shift $V_u$ measured as a function of ${\bf e}_{\rm k} \cdot {\bf e}_{\rm B}$, where we vary  ${\bf e}_{\rm B}={\bf B}_{\rm tot}/|{\bf B}_{\rm tot}|$ in respect  to ${\bf e}_{\rm k}$  by adding an extra magnetic field $B_{\rm ex}{\bf e}_{\rm k}$ to a bias magnetic field ${\bf B}_{\rm{bias}}=B_0 {\bf e}_x$, i.e., ${\bf B}_{\rm tot}={\bf B}_{\rm{bias}}+B_{\rm ex}{\bf e}_{\rm k}$.
The red line shows the result of a linear fitting to the data and indicates the degree of ellipticity of lattice laser polarization $\xi_L = 2.6(7) \times 10^{-4}$, where we use $|\tilde \kappa^{v}|/h = 0.22(5)$~Hz in Ref.~\cite{Westergaard2011}. 


\section*{Tensor light shift}
\begin{figure}[t]
\includegraphics[width = \linewidth, bb=30 400 530 750]{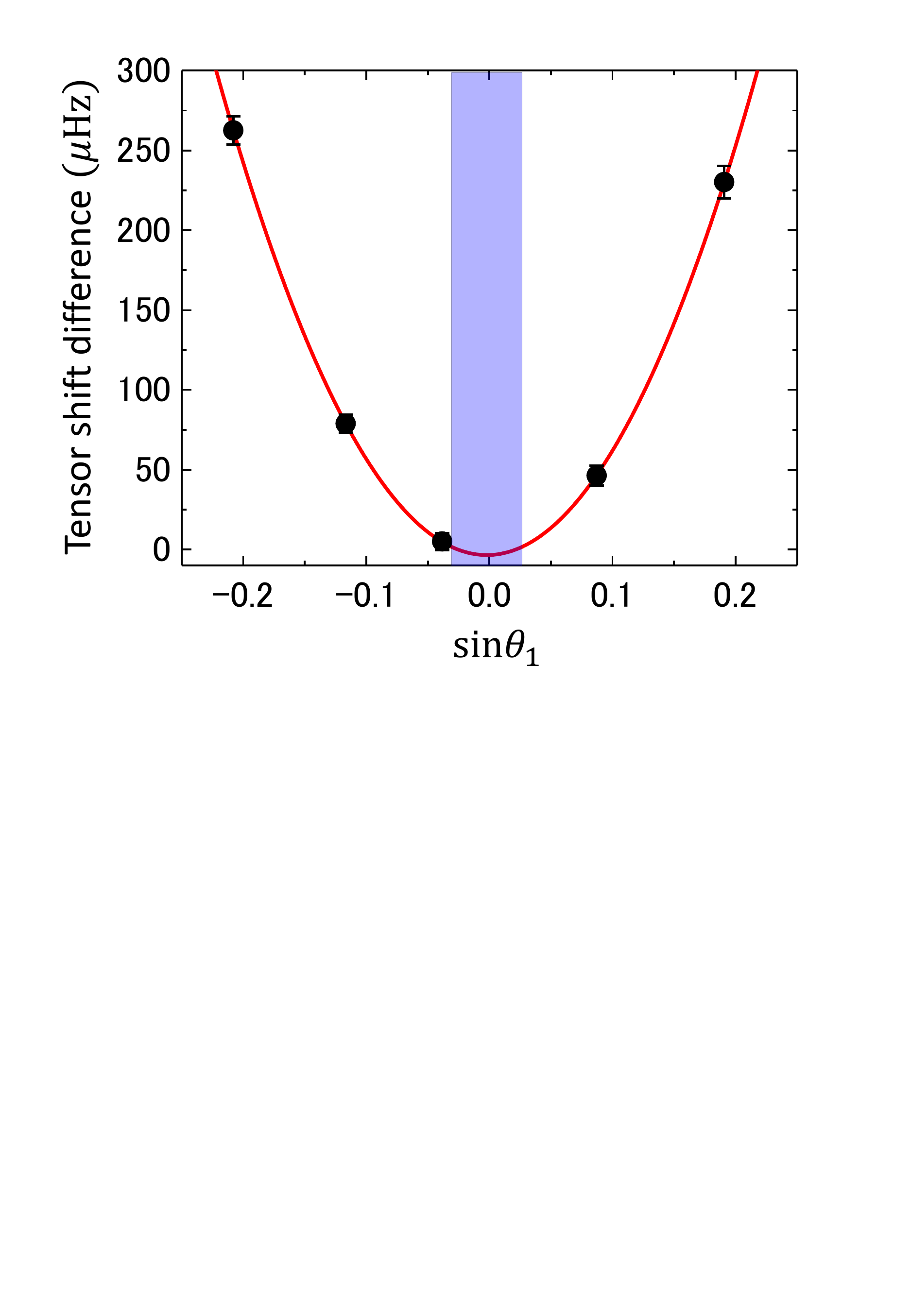}
  \caption{The dependence of the tensor light shift difference on the angle $\theta_1 $ of the quantization axis.
  The red line shows the result of a parabolic fitting with $\sin\theta_1$.}
  \label{tensor}
\end{figure}

We measure the clock transition frequency difference for the $m_F=9/2$ substate,  $\Delta \nu_{\rm tens}({u},\theta_1,\theta_0)=\nu ({u},\theta_1)-\nu({u},\theta_0)$ with the angle $\theta_j$  given by $|{\bf e_{\epsilon}} \cdot {\bf e}_{{\rm B}_j}|^2=\cos^{2}\theta_j$.
We set $\theta_1$ and $\theta_0$ with and without adding an external magnetic field  $B_{\rm ex}{\bf e}_{\rm k}$, where $|\theta_0|\ll1$ holds in the latter case.
Filled circles in Fig.~\ref{tensor} show the  tensor light shift difference   $\Delta \nu_{\rm tens}(\bar{u},\theta_1,\theta_0)/\bar{u}$ measured for $\bar{u}=1050$.
This tensor light shift difference is given by  $\Delta T_u(\theta_1,\theta_0,m_F)=T_u(\theta_1,m_F)-T_u(\theta_0,m_F)$. 
For $|\theta_0|\ll1$,  the equation is approximated by
\begin{equation}
\Delta T_u(\theta_1,m_F)= -3[3m_{F}^{2}-F(F+1)]\tilde \kappa ^{t} \sin^2\theta_1. 
\end{equation}
For $m_F=9/2$, this corresponds to
\begin{equation}
\Delta T_u(\theta_1)=-108~\tilde \kappa ^{t}  \sin^2\theta_1.
\end{equation} 
The red line shows the fit to the data, which gives $\tilde \kappa ^{t}/h = -58.1(2.9)~\mu $Hz and is consistent with the result of Ref.~\cite{Westergaard2011}.

The orientation of the bias magnetic field ${\bf e}_{\rm B}$ is kept inside the blue shaded area in Fig.~\ref{tensor} 
for our typical operation conditions.
The angle uncertainty can be introduced by the environmental magnetic field fluctuation 
of about 2~$\mu \rm{T}$ in arbitrary directions, which corresponds to $|\sin\theta_1|<0.029$.
Therefore, we estimate  the fractional uncertainty of the tensor light shift $\Delta T_u(\theta) \Bar{u}^{\rm op}/\nu_0<8.7\times 10^{-19}$  for  $\Bar{u}^{\rm op}=72$.

\section*{Density shift}
\begin{figure}[t]
\includegraphics[width = \linewidth, bb=30 400 530 750]{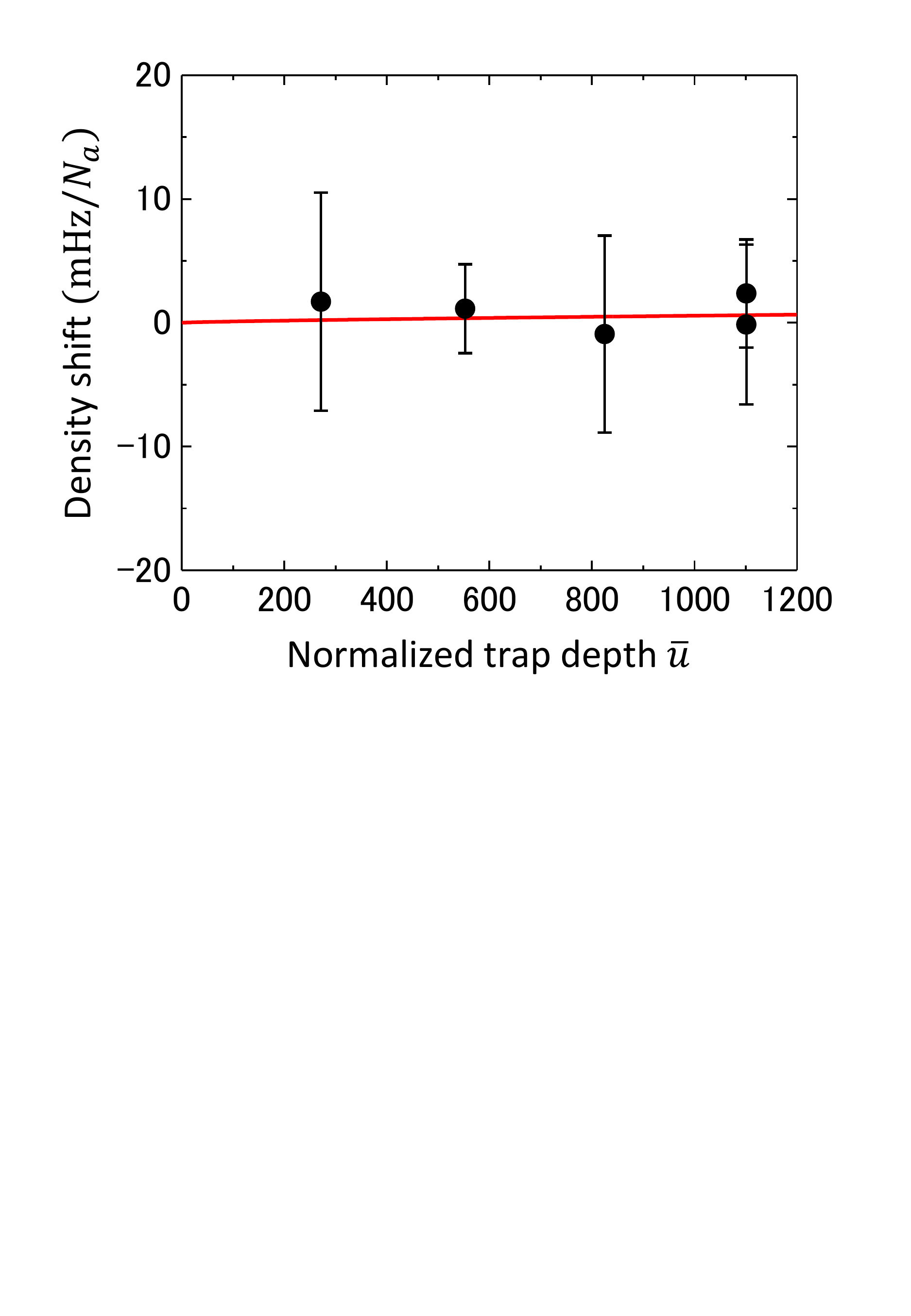}
  \caption{Density shift as a function of effective trap depth.  
}
  \label{fig:collision}
\end{figure}

The density shift in  Fig. \ref{fig:collision} is measured by varying the atom number $N$ for each trap  depth $\Bar{u}$.
For the adiabatic ramp of the lattice trap, density shift will scale as $N\Bar{u}^{3/4}$~\cite{Swa12} as shown by the red curve.
As we moderate the density shift by setting a large beam diameter $w\approx 60\, \mu$m for the 1D lattice, the density shift corrections are about  $0.6(9)~\rm{mHz}$ even at $\Bar{u}=1100$.

\section*{Approximate expressions for the operational magic intensity and frequency}
We derive approximate formulae to derive  operational magic intensity ($\bar{u}^{\rm{op}}$) and frequency ($\delta_L^{\rm{op}}$) determined by (i)
$\nu_{\rm LS}(u,\delta_L,0)=0$ and ${\partial \nu_{\rm LS}(u,\delta_L,0)}/{\partial u}=0$.
For low intensities of $(0<)u<100$, which include $\bar{u}^{\rm{op}}=72$, and small detuning $\delta_L^{\rm op}$, the light shift model in Eq.~(1) in the main text can be simplified by neglecting  terms with small contributions as 
\begin{align}
&h \nu_{\rm LS}\approx 
-\tilde \alpha^{qm} \left(n+\textstyle{1 \over 2} \right) u^{1/2}  - \frac{\partial \tilde \alpha^{\rm E1}}{\partial \nu} \delta_L u - \tilde \beta \, u^2,\\
&h \frac{\partial \nu_{\rm LS}}{\partial u} \approx -\textstyle{1 \over 2} \tilde \alpha^{qm} \left(n+\textstyle{1 \over 2} \right) u^{-1/2}
- \frac{\partial \tilde \alpha^{\rm E1}}{\partial \nu} \delta_L - 2 \tilde \beta \, u. 
\label{eq:LS2}
\end{align}
Solving the simultaneous equations with $n=0$, we find the operational magic intensity and frequency to be
\begin{align}
&\bar{u}^{\rm{op(i)}} = \left( \frac{\tilde \alpha^{qm}}{4\tilde \beta} \right)^{2/3}\\
&\delta_L^{\rm{op(i)}} = -{3 \tilde \beta}\left({\frac{\partial \tilde \alpha^{\rm E1}}{\partial \nu}} \right)^{-1}
\left( \frac{\tilde \alpha^{qm}}{4 \tilde \beta}  \right)^{2/3}.
\end{align}
The result conveniently  approximates operational conditions that allows cancelling out the light shift.

Similarly, the second derivative of Eq.~(12) is given by
\begin{equation}
h \frac{\partial^2 \nu_{\rm LS}}{\partial u^2} \approx \textstyle{1 \over 4} \tilde \alpha^{qm} \left(n+\textstyle{1 \over 2} \right) u^{-3/2} - 2 \tilde \beta. 
\end{equation}
The other option for the operational condition to operate at (ii) the inflection point $({{\textstyle{{\partial^2 {\nu _{{\rm{LS}}}}} \over {\partial u^2}} }}|_{u=\bar{u}^{\rm op}} = 0)$ with $n=0$ is, therefore, given by 
\begin{equation}
\bar{u}^{\rm{op(ii)}} = \left( \frac{\tilde \alpha^{qm}}{16\tilde \beta} \right)^{2/3}.
\end{equation}
By imposing the condition ${\partial \nu_{\rm LS}(u,\delta_L,0)}/{\partial u}=0$ in Eq.~(13), we obtain the operational frequency as,
\begin{equation}
\delta_L^{\rm{op(ii)}} =\left({\frac{\partial \tilde \alpha^{\rm E1}}{\partial \nu}} \right)^{-1}
\left(-\textstyle{1 \over 4}\tilde \alpha^{qm} u^{-1/2} -2 \tilde \beta u  \right).
\end{equation}

In both cases (i) and (ii), it is essential to have the same sign for $\tilde \alpha^{qm}$ and $\tilde \beta$ for these operational depths to exist.

\section*{Summary of our results}
Table~\ref{tab1:summary} summarizes the light-shifts  parameters  determined in this work. 

\begin{table}[h]
\caption{\label{tab1:summary}
Light shift parameters for Sr determined in our experiment.}
\begin{center}
\scalebox{0.9}{
\begin{tabular}{l c}
\hline\hline
Parameters  & Our results \\
\hline
 $\alpha^{\rm E1}/h$ 	& $54.1(4.3)\, \rm{kHz/(kW/cm}^2) $ \\ 
 $(1/h) \partial \tilde \alpha^{\rm E1}/\partial \nu_L $			& $1.735(13)\times10^{-11} $ \\ 
 $\nu^{\rm E1} $					& $368~554~465.1(1.0)\,  \rm{MHz} $ \\ 
 $\tilde \beta/h $			& $-0.461(14) \, \mu \rm{Hz} $\\ 
 $\tilde \alpha^{qm}/h  $		& $-0.962(40)\, \rm{mHz} $ \\ 
\hline\hline
\end{tabular}
}
\end{center}
\end{table}
